\documentstyle[prc,aps,epsfig,multicol]{revtex}
\draft
\tightenlines
\begin{document}
\title{Multistrange baryon production in relativistic heavy ion 
collisions}
\author{Subrata Pal, C. M. Ko, and Zi-wei Lin}
\address{Cyclotron Institute and Physics Department,
Texas A\&M University, College Station, Texas 77843-3366}

\maketitle

\begin{abstract}
Using a multiphase transport model, we study the production of
multistrange baryons from the hadronic matter formed in relativistic 
heavy ion collisions. The mechanism we introduce is the 
strangeness-exchange reactions between antikaons and
hyperons. We find that these reactions contribute significantly to
the production of multistrange baryons in heavy ion collisions at 
SPS energies, which has been found to be appreciably enhanced. 
We have also made predictions for multistrange baryon production 
in heavy ion collisions at RHIC and found a similar enhancement.

\medskip
\noindent PACS numbers: 25.75.-q, 24.10.Lx
\end{abstract}

\begin{multicols}{2}

\section{introduction}

One possible signal for the quark-gluon plasma (QGP) that is expected to 
be formed in relativistic heavy ion collisions is enhanced production
of strange particles \cite{rafelski}, particularly those consisting 
of multistrange quarks such as cascade and omega as well as their 
antiparticles. The argument is that the rate for strange hadron 
production is small in hadronic matter due to the large threshold 
and small cross sections, which is contrary to the large production
rate for strange quarks in the quark-gluon plasma \cite{koch}.
For the lightest strange hadrons such as kaons and antikaons, experiments 
have shown enhanced production in heavy ion collisions at all energies 
available from SIS ($\sim 1A$ GeV) \cite{sis}, AGS ($\sim 10A$ GeV) 
\cite{ags1}, and SPS ($\sim 158A$ GeV) \cite{sps}. Studies based on 
transport models show that this enhancement can be explained by hadronic 
scattering alone \cite{transport}. However, these models have failed in 
accounting for the enhanced production of multistrange baryons ($\Xi$, 
$\Omega$) by about a factor of three, and the discrepancy is even larger 
for their antiparticles \cite{urqmd}.  To increase the production of these 
particles requires either to increase the string tension or to decrease
the constituent quark mass in the fragmentation of the  
initial strings in the dense matter \cite{urqmd}. It has also been
suggested that the observed enhancement of multistrange baryons and
antibaryons may be due to topological defects arising from the
formation of disoriented chiral condensate in the initial high
density stage of collisions \cite{kapusta}. Of course, the enhancement
could simply be a result of the formation of the quark-gluon plasma
during collisions.  However, to establish strangeness enhancement 
as a signal for the quark-gluon plasma, we need to exclude any conventional 
mechanisms. As recently suggested by Vance \cite{vance}, 
strangeness-exchange reactions between antikaons and hyperons as well 
as between their antiparticles can contribute significantly to the 
production of multistrange baryons and antibaryons in relativistic 
heavy ion collisions.

In this paper, we shall determine the contribution to multistrange
baryon production from the hadronic matter in heavy ion collisions 
at both SPS and RHIC energies, using a recently developed multiphase 
transport (AMPT) model \cite{ampt1}.  In Section \ref{transport}, 
we describe briefly the multiphase transport model. The 
strangeness-exchange reactions are then discussed in Section \ref{strange}.
Results from the AMPT model are given in Section \ref{cern} for heavy 
ion collisions at SPS and in Section \ref{rhic} for heavy ion 
collisions at RHIC. Finally, a summary is given in Section \ref{summary}.

\section{multiphase transport model}\label{transport}

In the AMPT model, the initial conditions are obtained from the HIJING 
model \cite{hijing} by using a Woods-Saxon radial shape for the colliding 
nuclei and including the nuclear shadowing effect on minijet partons
via the gluon recombination mechanism of Mueller-Qiu \cite{amueller}. 
After the colliding nuclei pass through each other, the Gyulassy-Wang model
\cite{gyulassy} is then used to generate the initial space-time information 
of partons. Subsequent time evolution of the parton phase-space 
distribution is modeled by Zhang's Parton Cascade (ZPC) \cite{zpc},
which at present includes only the gluon elastic scattering.
After minijet partons stop interacting, they combine with their parent strings 
and are then converted to hadrons using the Lund string fragmentation 
model \cite{lund,jetset} after an average proper formation time of 0.7 fm/c. 
Dynamics of the resulting hadronic matter is described by a relativistic 
transport model (ART) \cite{art}.

The parameters in the AMPT model are fixed using the experimental 
data from central Pb+Pb collisions at center of mass energy of $17A$ GeV 
\cite{na49}. Specifically, we have included in the Lund string fragmentation 
model the popcorn mechanism for baryon-antibaryon production in order to 
describe the measured net baryon rapidity distribution. Also, to account 
for the pion and enhanced kaon yields, we have modified the values of 
the parameters in the string fragmentation function.  The same parameters 
are then used to study heavy ion collisions at RHIC energies. 
The model is able to describe the PHOBOS data \cite{phobos} for the 
pseudorapidity distribution of charged particles \cite{ampt2}. 
Also, the predicted $\bar p/p$ and $K^-/\pi^-$ ratios are consistent with 
the data from the STAR \cite{star} and BRAHMS \cite{brahms} Collaborations.
Furthermore, it gives an elliptic flow \cite{lin} which is comparable to 
that measured in the STAR experiment \cite{v2}.

\section{strangeness-exchange reactions}\label{strange}

To include multistrange baryon and antibaryon production, we consider
the following reactions:
\begin{eqnarray}\label{mstrange}
{\bar K}\Lambda\leftrightarrow\Xi\pi,~~
{\bar K}\Sigma\leftrightarrow\Xi\pi,~~{\rm and}~~
{\bar K}\Xi\leftrightarrow\Omega\pi.
\end{eqnarray}

\begin{figure}[ht]
\centerline{\epsfig{file=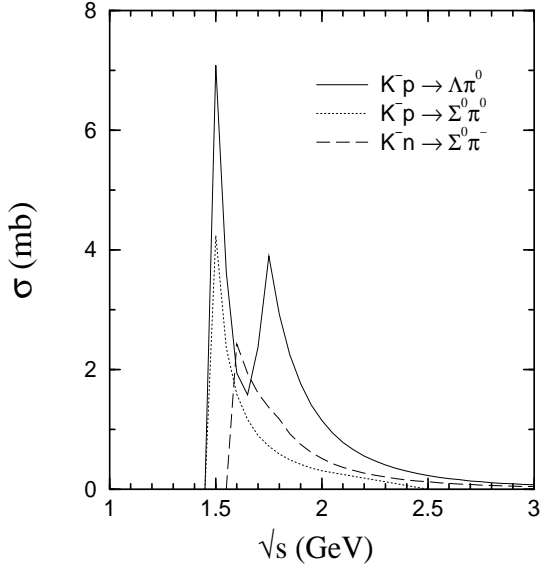,width=2.8in,height=3.0in,angle=0}}
\caption{Experimental cross sections for $K^-p\to\Lambda\pi^0$,
$K^-p\to\Sigma^0\pi^0$ and $K^-n\to\Sigma^0\pi^-$ as a function of 
center of mass energy of the interacting antikaon and nucleon.}
\label{cross}
\end{figure}

Since there is no experimental information on these cross sections,
we assume that they are the same as the cross section for
$\bar KN\to\Sigma\pi$, which is the isospin averaged cross
section for converting a nucleon to a sigma and can be related to the 
cross sections for $K^-p\to\Sigma^0\pi^0$, and $K^-n\to\Sigma^0\pi^-$ by
\begin{eqnarray}\label{kn}
\sigma_{\bar KN\to\Sigma\pi}&=&\frac{3}{2}(\sigma_{K^-p\to\Sigma^0\pi^0}
+\sigma_{K^-n\to\Sigma^0\pi^-}).
\end{eqnarray}

The cross sections on the right hand side are known empirically and
have been parameterized in Ref. \cite{cugnon} as follows,
\begin{eqnarray}
\sigma_{K^-p \to \Sigma^0\pi^0}=0.6p^{-1.8}\ [{\rm mb}];\ 0.2
\leq p \leq 1.5\ {\rm GeV/c},
\end{eqnarray}
\begin{eqnarray}
\sigma_{K^-n \to \Sigma^0\pi^-}= \left\{ \begin{array}{l}
1.2p^{-1.3} \ [{\rm mb}], \\
~~~~~~~~~\mbox{${\rm if} \ 0.5 \leq p \leq 1\ {\rm GeV/c}$;}\\
1.2p^{-2.3}\ [{\rm mb}], \\
~~~~~~~~~\mbox{${\rm if} \ 1 \leq p \leq 6\ {\rm GeV/c}$.}  
\end{array}\right.
\end{eqnarray}
In the above, $p$ is the $K^-$ momentum in the laboratory frame.
These cross sections are shown in Fig. \ref{cross} as functions of
center of mass energy of the interacting antikaon and nucleon.

Empirically, the cross section for $K^-p\to\Lambda\pi^0$ is also known
and has been parameterized \cite{cugnon} as
\begin{eqnarray}\label{knlp}
\sigma_{K^-p\to\Lambda\pi^0}= \left\{ \begin{array}{l}
50p^2-67p+24\ [{\rm mb}], \\
~~~~~~~~~\mbox{if \ $0.2 \leq p \leq 0.9\ {\rm GeV/c}$;} \\
3p^{-2.6}\ [{\rm mb}], \\
~~~~~~~~~\mbox{if \ $0.9 \leq p \leq 10\ {\rm GeV/c}$,}
\end{array}\right.
\end{eqnarray}

As shown in Fig. \ref{cross}, $\sigma_{K^-p\to\Lambda\pi^0}$ has a 
similar magnitude as $\sigma_{K^-p\to\Sigma^0\pi^0}$ and
$\sigma_{K^-n\to\Sigma^0\pi^-}$ except that it has a resonance contribution
at $\sqrt{s}\sim 1.8$ GeV. It may be noted that the channel 
$K^-p\to\Lambda\pi^0$ has resonances while $\Omega$ has no resonance
and the branching ratios of the decay channels for the resonances 
of $\Xi$ are not well determined. 
Therefore for a conservative estimate we assume that the reactions in 
Eq. (\ref{mstrange}) are similar to $\bar K N\to\Sigma\pi$, which does 
not show any resonance structure.

We believe the assumption that all strangeness-exchange reaction
cross sections have similar magnitudes is reasonable
as these reactions are similar except for differences in their
particle masses. Justification of this assumption is being studied
in an effective hadronic model \cite{chli} that is based on 
SU(3) flavor symmetry. Preliminary results indeed show that 
the cross sections for these reactions have comparable values.

\section{multistrange baryon production from heavy ion collisions}

\subsection{Perturbative approach}\label{perturbative}

Since the abundance of multistrange baryons is small in heavy ion 
collisions, they can be treated perturbatively in the AMPT model 
as in kaon production from heavy ion collisions at low energies
\cite{pert}, where the collision dynamics is assumed 
not to be affected by the production of these particles.

First, we consider the process ${\bar K}\Lambda\to\Xi\pi$,
in which both $\bar K$ and $\Lambda$ are treated explicitly, i.e., 
nonperturbatively, in the AMPT model. The $\bar K\Lambda$ collision
is treated by taking their total scattering cross section to be 
$\sigma_{\bar K\Lambda}=20$ mb. $\bar K$ and $\Lambda$ then make
a collision if their impact parameter is less than 
$\sqrt{\sigma_{\bar K\Lambda}/\pi}$. When this occurs,  
a $\Xi$ is produced with a probability given by 
$P_\Xi=\sigma_{\bar K\Lambda\to\Xi\pi}/\sigma_{\bar K\Lambda}$, 
which is stored as the $\Xi$ formation probability. At the same time,
both $\bar K$ and $\Lambda$ have a probability of $P_{\Xi}$ being 
destroyed, and a pion is also produced with the same probability. 
Otherwise, the momenta of $\bar K$ and $\Lambda$ are changed according 
to an isotropic distribution.

To treat the inverse reaction of $\Xi$ absorption by a pion, i.e,
$\Xi\pi\to\bar K\Lambda$, we use the cross section determined by
detailed balance from Eq. (\ref{kn}). If a collision occurs, the $\Xi$ 
is destroyed. The pion has a destruction probability given by the 
$\Xi$ formation probability $P_\Xi$ and simultaneously a $\bar K$ and 
$\Lambda$ are produced with an isotropic momentum distribution.

The above perturbative approach can be similarly used for the reaction
$\bar K\Sigma\leftrightarrow\Xi\pi$. However, it needs to be modified for the
reaction $\bar K\Xi\leftrightarrow\Omega\pi$ as both $\Xi$ and $\Omega$
are treated perturbatively. Again, we take the $\bar K\Xi$ 
scattering cross section to be $\sigma_{\bar K\Xi}=20$ mb. Whenever
they collide, the $\Xi$ have a destruction probability 
given by $\sigma_{\bar K\Xi\to\Omega\pi}/\sigma_{\bar K\Xi}$ 
while an $\Omega$ is produced with a probability given by the above
probability multiplied by the $\Xi$ formation probability $P_\Xi$,
which is then stored as the $\Omega$ formation probability $P_\Omega$.
The destruction of $\bar K$ and simultaneous production of $\pi$ 
has a probability of $P_\Omega$. The inverse reaction 
$\Omega\pi\to\bar K\Xi$ can be similarly treated as for 
$\Xi\pi\to\bar K\Lambda$.

\subsection{Multistrange baryon production at SPS}\label{cern}

\begin{figure}[ht]
\centerline{\epsfig{file=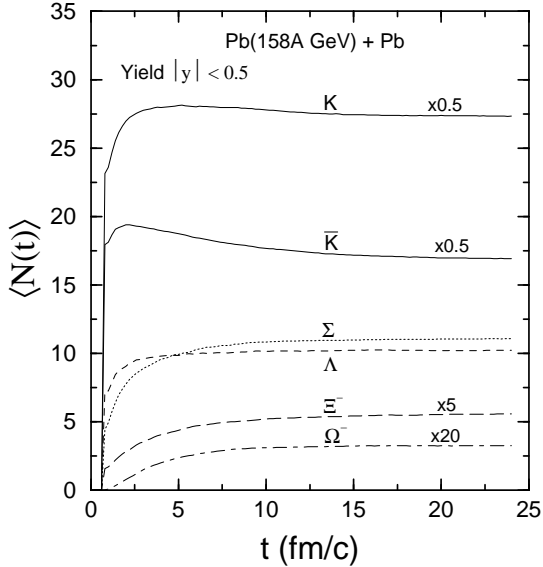,width=2.8in,height=3.0in,angle=0}}
\vspace{0.3cm}
\caption{Time evolution of midrapidity hadrons for Pb+Pb collisions 
at SPS energy of $\sqrt s = 17A$ GeV at an impact parameter of 
$b\leq 3$ fm in the AMPT model.}
\label{yield17}
\end{figure}

In Fig. \ref{yield17}, we show the time evolution of the abundance of
midrapidity kaons (including $K^*$), antikaons (including $\bar K^*$), 
$\Lambda$, $\Sigma$, $\Xi^-$, and 
$\Omega^-$ in Pb+Pb collisions at $\sqrt s=17A$ GeV at an impact 
parameter $b=0-3$ fm. As is evident from the figure, most multistrange 
baryons are produced within 10 fm/c after the initial 
contact of the colliding nuclei when the energy density is high. 
Compared to the initial yield obtained from HIJING, the dynamical evolution
of the system leads to about $30\%$ and $70\%$ increase in the 
$\Lambda$ and $\Xi^-$ production, while most of the $\Omega^-$s are
produced from hadronic rescattering. Therefore the observed enhancement of 
multistrange baryons \cite{sps} can be largely accounted by the 
strangeness-exchange reactions in the hadronic matter. 
Furthermore, the final yield of strange baryons 
in the AMPT model are in reasonable agreement with that 
obtained by the WA97 Collaboration \cite{sps}. The $\Omega^-$ production
in our model is however a factor of two smaller compared to the data.
It may be mentioned that using a rate equation approach, it was 
demonstrated \cite{carsten} that multimesonic reactions 
$\bar Y + N \leftrightarrow n\pi + n_YK$ may lead to an enhanced antistrange 
hyperon $\bar Y$ production in a purely hadronic scenario. In the present 
transport model we have neglected these multiparticle reactions which 
in conjunction with the strangeness-exchange reactions could lead to a better
reproduction of the experimental data for multistrange baryons.
Alternate to the multiparticle reaction, as for example, 
$3\pi + 2\bar K \leftrightarrow \Xi + \bar N$, the process
$\bar K(\bar K^*) + \bar K(\bar K^*) \leftrightarrow \Xi + \bar N$
may be used instead in the transport model to generate $\Xi$.

\begin{figure}[ht]
\centerline{\epsfig{file=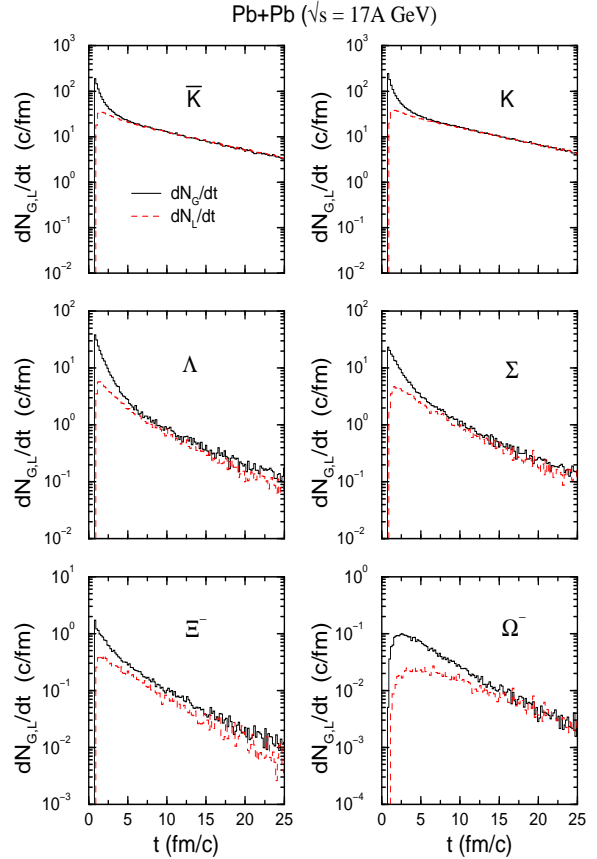,width=3.0in,height=4.5in,angle=0}}
\vspace{0.3cm}
\caption{Production and absorption rates of strange particles
in heavy ion collisions at SPS as functions of time.}
\label{loga17}
\end{figure}

To see if strange hadrons reach chemical equilibrium in the collisions,
we show in Fig. \ref{loga17} the time evolution of their production 
(solid lines) and absorption rates (dashed lines).
It is seen that most strange particles such as the hyperons
and especially the kaons and antikaons approach equilibrium 
as their production and absorption rates become comparable before the 
system freezes out. So the abundance of 
these particles in heavy ion collisions at SPS are consistent with 
that given by the equilibrium models \cite{braun,hamieh}.

\begin{figure}[ht]
\centerline{\epsfig{file=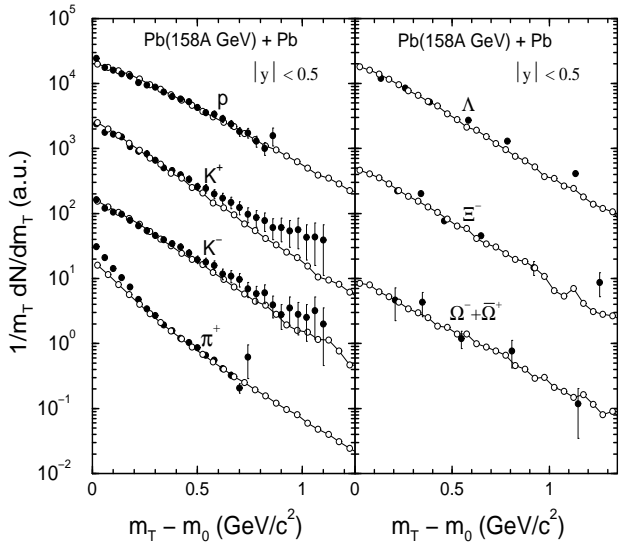,width=3.2in,height=2.8in,angle=0}}
\vspace{0.3cm}
\caption{Transverse mass spectra for midrapidity protons, kaons, antikaons,
pions (left panel), and lambda, cascade, omega (right panel) in $158A$ GeV 
Pb+Pb central collisions. The solid circles are the experimental data
for $\sim 5\%$ most central collisions from the NA44 Collaboration [31]
(left panel), and $\sim 4\%$ central events from the
WA97 Collaboration [32] (right panel). The open circles are the 
AMPT model calculations for impact parameter of $b\le 3$ fm.} 
\label{mt17}
\end{figure}

In Fig. \ref{mt17} we depict by open circles the transverse mass spectra 
of hadrons at midrapidity obtained in the AMPT model for the SPS energy. 
These results are 
obtained by using mean-field potentials for strange baryons that 
are related to the nucleon potential according to their light quark 
contents. Specifically, relative to the nucleon potential, which is 
taken to be a stiff Skyrme potential with compressibility of 380 MeV,  
the lambda and sigma potentials are 2/3, the cascade potential is 1/3, 
and the omega potential is zero. We see that the theoretical transverse 
mass spectra of the hadrons agree well with the experimental data 
shown by solid circles \cite{NA44,Antin}. The kaon transverse mass 
spectrum is however somewhat steeper than the experimental one. 
This may be due to the weak kaon mean-field potential at high density 
that is based on the impulse approximation \cite{art} and is used in the 
present study.

\begin{figure}[ht]
\centerline{\epsfig{file=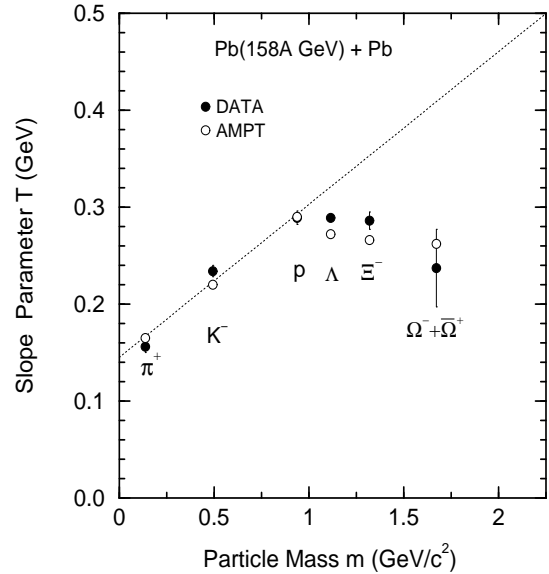,width=2.8in,height=3.in,angle=0}}
\vspace{0.3cm}
\caption{Inverse slope parameters of various particles in heavy 
ion collisions at SPS. The open circles are the AMPT calculations
and the solid circles are the experimental data [32]. The dotted line
represents a linear parameterization (see text). }
\label{slope17}
\end{figure}

The transverse mass spectra can be approximately fitted by an exponential 
function $exp(-m_T/T)$, where $T$ is the inverse slope parameter. 
The slope parameter may provide useful information on the thermal
motion and the strong collective transverse flow observed at SPS
energy. In Fig. \ref{slope17}, we show the inverse slope parameter of these 
particles as calculated from the AMPT model (open circles) and compare 
them with those extracted from the experimental data (solid circles)
\cite{Antin}. The agreement between theory and experiment is reasonable. 
The linear increase of the inverse slope parameter from pion to kaon and 
to proton is indicative of the development of collective transverse flow in 
the system. For nonstrange hadrons, $T$ may be parametrized by 
$T=T_{\rm fo} + m\langle\beta_t\rangle^2$ (dotted line), where the 
freeze-out temperature $T_{\rm fo}\approx 145$ MeV and average transverse 
flow $\langle\beta_t\rangle \simeq 0.39$c.
As in the experimental data, the inverse slope parameters for strange 
baryons in the AMPT model are smaller than that for the proton.
This is due to both the fact that strange hadrons are produced earlier
in the collision when the energy density is high and have smaller
scattering cross sections with pions and nucleons than nonstrange hadrons
and thus freeze out quite early \cite{Hecke}. Note that for proton
with slope parameter $T=0.29$ GeV, the contribution from HIJING is about 
0.16 GeV, while the mean-field causes a further increase by 
$\approx 0.03$ GeV.

\subsection{Multistrange baryon production at RHIC}\label{rhic}

\begin{figure}[ht]
\centerline{\epsfig{file=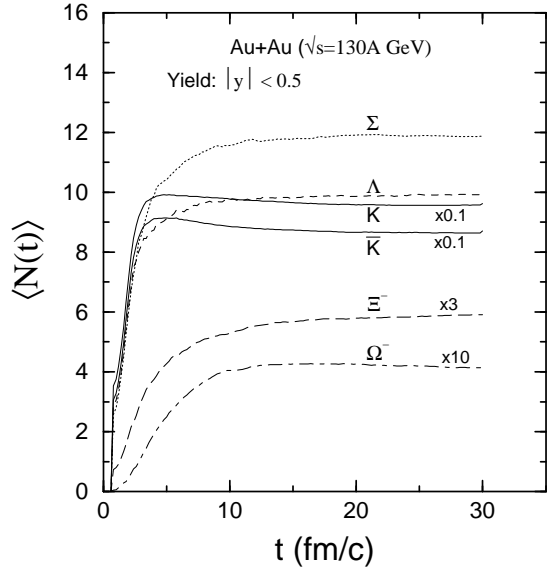,width=2.8in,height=3.0in,angle=0}}
\vspace{0.3cm}
\caption{Time evolution of midrapidity hadrons for Au+Au collisions 
at RHIC energy of $\sqrt s = 130A$ GeV at an impact parameter of 
$b\leq 3$ fm in the AMPT model.}
\label{yield130}
\end{figure}

The predictions of the AMPT model for multistrange baryon production
in Au+Au collisions at RHIC energy of $\sqrt s=130A$ GeV are given in 
Fig. \ref{yield130} for their time evolution. Compared to that from 
heavy ion collisions at SPS, we see that the abundance of the 
multistrange particles at this RHIC energy is increased by about a 
factor of two. We have also examined the time evolution of the production
and absorption rates of strange particles. It is found that both
kaons and antikaons approach chemical equilibrium. On the other hand, 
strange baryons such as $\Sigma$ and $\Xi$ are somewhat out of 
equilibrium as their production rates are larger than their absorption 
rates at freeze out. Again, including strange baryon production from 
multiparticle reactions would lead to chemical equilibrium for
these particles as well. Because of its larger absorption cross section 
by the pion compared to other strange baryons, the $\Omega$ does reach 
equilibrium during collision. 

We note that our predicted ratios of multistrange 
baryons to mesons, such as $\Xi^-/K^-$ and $\Omega^-/\pi^-$,
are similar to that predicted by the thermal model \cite{braun1}.
In such statistical model analysis of the experimental data at the RHIC energy 
of $\sqrt s=130A$ GeV, the hadron yields are in thermal and 
chemical equilibrium with a baryon chemical potential of $\mu_B \simeq 51$ 
MeV and freeze-out temperature of $T \simeq 175$ MeV. 
The strangeness-exchange reactions introduced in the present study
are thus another mechanism for multistrange baryons to achieve chemical 
equilibrium in relativistic heavy ion collisions.

\begin{figure}[ht]
\centerline{\epsfig{file=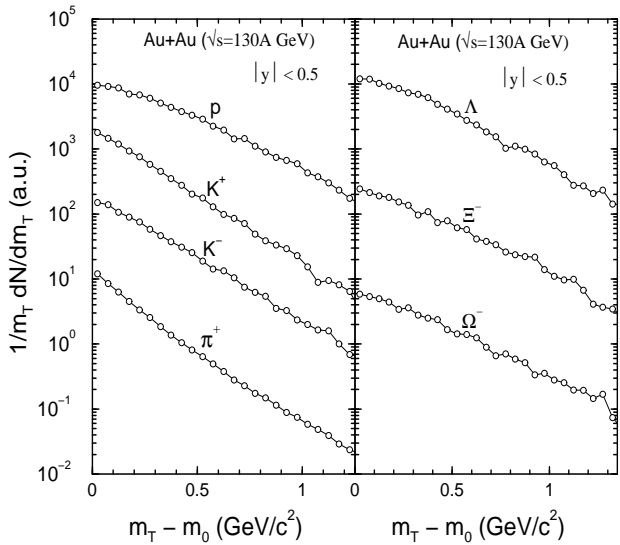,width=3.2in,height=2.8in,angle=0}}
\vspace{0.3cm}
\caption{Same as Fig. \ref{mt17} but for Au+Au collisions at 
$\sqrt s = 130A$ GeV for impact parameter of $b\le 3$ fm in the AMPT model.} 
\label{mt130}
\end{figure}

\begin{figure}[ht]
\centerline{\epsfig{file=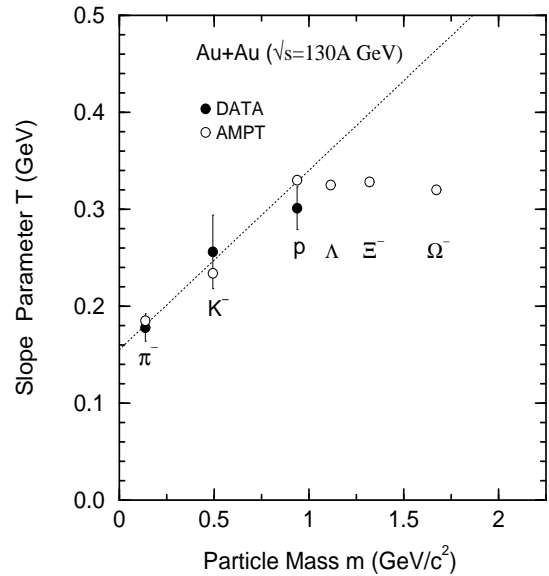,width=2.8in,height=3.in,angle=0}}
\vspace{0.3cm}
\caption{Inverse slope parameters of various particles in heavy 
ion collisions at RHIC. The open circles are the AMPT calculations
and the solid circles are the experimental data [35]. The dotted line
represents a linear parameterization (see text). }
\label{slope130}
\end{figure}

In Fig. \ref{mt130} we show the transverse mass spectra of midrapidity
hadrons for central Au+Au collision at the RHIC energy $\sqrt s = 130A$ GeV.
The spectra for protons, pions and kaons are consistent with the
preliminary data from the PHENIX Collaboration \cite{phenix}. The agreement
with this data is clearly evident in Fig. \ref{slope130} where the slope
parameters obtained by fitting the spectra in the same $p_t$ range with
the exponential function is depicted. The mass dependence of $T$ is 
qualitatively similar to that at SPS energy. At the RHIC energy the slope
parameter from pion to proton also exhibits a strong dependence on the 
particle mass with the freeze-out temperature 
$T_{\rm fo} \approx 155$ MeV and the average collective velocity 
$\langle \beta_t\rangle \approx 0.43$c. As at SPS the slope
parameter for strange and multistrange baryons reveals a plateau since these
particles, mostly generated by strangeness-exchange reactions, are weakly 
interacting and decouples rather early from the system.

\section{summary}\label{summary}

In this paper, we have used a multiphase transport model, that includes
both an initial partonic matter and a final hadronic matter, to study the
production of multistrange baryons from the 
strangeness-exchange reactions in the hadronic matter. The cross sections
for these reactions are assumed to be the same as the empirically
known cross sections for antikaon-nucleon to hyperon-pion.
For heavy ion collisions at SPS energies, we find that these reactions
lead to an enhanced production of multistrange particles, comparable
to that observed by the WA97 collaboration. For heavy ion collisions
at RHIC, a similar enhancement is obtained from our model.
We further find that
the slope parameters for the (multi-)strange baryons exhibit a plateau as a
function of particle mass at both SPS and RHIC energies.

We have not considered in the present study the hadronic contribution to 
multistrange antibaryons. For such a study, we need to extend the AMPT 
model to treat antibaryons similar to that for baryons, which is
currently under way.

\section*{acknowledgment}

This work was supported by the National Science Foundation under Grant 
No. PHY-9870038, the Welch Foundation under Grant No. A-1358, and the 
Texas Advanced Research Program under Grant No. FY99-010366-0081.

\end{multicols}

\end{document}